\documentclass{Rinton-P9x6}

\def\ss{\scriptscriptstyle}
\def\ga{\mathrel{\raise.3ex\hbox{$>$\kern-.75em\lower1ex\hbox{$\sim$}}}}
\def\la{\mathrel{\raise.3ex\hbox{$<$\kern-.75em\lower1ex\hbox{$\sim$}}}}
\def\gev{{\rm \, Ge\kern-0.125em V}}
\def\tev{{\rm \, Te\kern-0.125em V}}
\def\gyr{{\rm \, G\kern-0.125em yr}}
\def\ohsq{\Omega_{\chi} h^2}

\def\nnl{\hfill\nonumber\\}

\def\slept{\widetilde \ell} 

\def\selR{{\widetilde e}_{\scriptscriptstyle\rm R}}

\def\stau{\widetilde \tau}
\def\snu{\widetilde \nu}
\def\stop{\widetilde t}

\def\m12{m_{1\!/2}}

\def\mz{m_{\ss Z}}

\def\gappeq{\mathrel{\rlap {\raise.5ex\hbox{$>$}}
{\lower.5ex\hbox{$\sim$}}}}
\def\lappeq{\mathrel{\rlap{\raise.5ex\hbox{$<$}}
{\lower.5ex\hbox{$\sim$}}}}

\begin{document}
\vskip -0.5in
{\large
\rightline{hep-ph/0307356}
\rightline{UMN--TH--2206/03}
\rightline{FTPI--MINN--03/17}
}
\vskip 0.5in
\begin{center}
{\LARGE{\bf The MSSM with Non-Universal Higgs Masses}~\footnote{To be published
in the proceeding of 
SUGRA20 conference, March 17 - 21, 2003, Northeastern University,
Boston, MA.}}
\end{center}
\begin{center}
\vskip 0.4in
{{\large \bf Yudi Santoso}}\\
\vskip 0.2in
{\it
{William I. Fine Theoretical Physics Institute,
University of Minnesota, Minneapolis, MN 55455, USA}}\\
\vskip 0.5in
{\large \bf Abstract}
\vskip 0.2in
\end{center}
I review a generalization of CMSSM/mSUGRA model in which the Higgs masses are
let to be free, namely the Non Universal Higgs Masses (NUHM) model. In our study
we explore this model by employing constraints from cosmology as well as from
particle physics to restrict the parameter space. We also calculate the
neutralino-proton cross section in the allowed regions.

\vfill

\newpage
\title{The MSSM with Non-Universal Higgs Masses}

\author{Yudi Santoso}

\address{William I. Fine Theoretical Physics Institute,
University of Minnesota, \\ Minneapolis, MN 55455, USA
\\E-mail: santoso@physics.umn.edu
}

\maketitle

\abstracts{ 
I review a generalization of CMSSM/mSUGRA model in which the Higgs masses are
let to be free, namely the Non Universal Higgs Masses (NUHM) model. In our study
we explore this model by employing constraints from cosmology as well as from
particle physics to restrict the parameter space. We also calculate the
neutralino-proton cross section in the allowed regions.}

\section{Introduction}  
In the Constrained Minimal Supersymmetric Standard Model (CMSSM) or  minimal
Supergravity model (mSUGRA) all soft breaking scalar masses are set to be
universal at the GUT scale. While the universality of the sfermion masses is
motivated by GUT theories and FCNC problems, universality in the Higgs sector
is less theoretically motivated. This inspires ones to generalize the CMSSM by
letting the Higgs masses, $m_1$ and $m_2$, be non-universal. One can then
choose the Higgs mixing parameter $\mu$ and the pseudoscalar Higgs mass $m_A$
as free parameters.  We call this model the Non Universal Higgs Masses (NUHM)
model. We have been studying this model in several
papers~\cite{NUHM} and find that it has some new features not seen in the
CMSSM. We also calculate the
neutralino-proton cross section for the allowed regions to find the possible
range of neutralino direct detection rate in this model.

\section{The NUHM Model}
The free parameters in the CMSSM are: the universal scalar mass $m_0$, the
universal gaugino mass $m_{1/2}$, the universal trilinear coupling $A_0$, all
three defined at the GUT scale, the ratio of the two Higgs vev $\tan \beta
\equiv v_2/v_1$, and ${\rm sign}(\mu)$ where $\mu$ is the Higgs mixing
parameter. The magnitude of $\mu$ is determined by the electroweak symmetry
breaking conditions
\begin{eqnarray}
m_A^2 & =& m_1^2 + m_2^2 + 2 \mu^2 + \Delta_A \nnl
\mu^2 &=& \frac{m_1^2 - m_2^2 \tan^2 \beta + \frac{1}{2} \mz^2 (1 - \tan^2 \beta)
+ \Delta_\mu^{(1)}}{\tan^2 \beta - 1 + \Delta_\mu^{(2)}} \, ,
\end{eqnarray}
where $\Delta_A$ and $\Delta_\mu^{(1,2)}$ are loop
corrections and $m_{1,2}(M_{GUT})=m_0$.  
In the NUHM $m_1$ and $m_2$ are no longer set equal to $m_0$ at the GUT
scale. Thus we can use $\mu$ and the pseudoscalar Higgs mass $m_A$ as our input 
parameters. 

Solving for $m^2_1$ and $m^2_2$ we can see that, if $m_A$ is too small
or $\mu$ is too large, then $m_1^2$ and/or $m_2^2$ can become negative and
large. This could lead to $m_1^2(M_{GUT}) + \mu^2(M_{GUT}) < 0$ and/or
$m_2^2(M_{GUT})
+ \mu^2(M_{GUT}) < 0$, which raises the question of vacuum stability as the
universe evolved from high temperature to current temperature. The requirement
that
electroweak symmetry breaking occurs far below the GUT scale forces us to
impose the conditions $m_{1,2}^2(M_{GUT})+ \mu^2(M_{GUT}) > 0$ as
extra constraints, which we call the GUT
stability constraints.\footnote{For a different point of view, however,
see~\cite{fors}.}

There are terms in the RGE, collectively called the $S$-term:   
\begin{eqnarray}
S &\equiv& \frac{g_1^2}{4} ( m_2^2 - m_1^2 +
        2 ( m_{\widetilde{Q}_L}^2  - m_{\widetilde{L}_L}^2 - 2
	m_{\widetilde{u}_R}^2 +
	m_{\widetilde{d}_R}^2 + m_{\widetilde{e}_R}^2 ) \nonumber \\ && \, + \, 
          ( m_{\widetilde{Q}_{3L}}^2 - m_{\widetilde{L}_{3L}}^2 - 2
	  m_{\widetilde{t}_R}^2 
	  + m_{\widetilde{b}_R}^2 + m_{\widetilde{\tau}_R}^2 )) \, , 
\label{defS}
\end{eqnarray}
which vanishes for universal model due to anomaly
cancellation. However it is nonzero for NUHM model.
Although we start with non-universality only in the Higgs sector, the sfermion
mass spectrum is also affected trough the interconnected RG equations.
The $S$-term appears, for example, in
the slepton masses running as follows (neglecting the Yukawas)
\begin{eqnarray}
 \frac{d m_{\widetilde{L}_L}^2}{dt} &=& \frac{1}{8 \pi^2} (-3 g_2^2 M_2^2 -
          g_1^2  M_1^2 - 2 S)  \nnl
 \frac{d m_{\widetilde{e}_R}^2}{dt} &=& \frac{1}{8 \pi^2} (-4 g_1^2 M_1^2 + 4 S)
 \, . 
\label{rges}
\end{eqnarray}
We can deduce immediately that if $S$ is large and negative we could have
$\widetilde{L}_L$ lighter that $\widetilde{e}_R$ which means that sneutrino
could be the
lightest supersymmetric particle (LSP).  

In summary the possible LSP in the NUHM are the lightest neutralino $\chi$, the
lightest stau $\stau_1$, right handed selectron $\selR$, sneutrinos $\snu$ and
the lightest stop $\stop_1$ (for
large $|A_0|$). However, sneutrino dark matter is mostly
excluded~\cite{Falk:1994es},  therefore we still assume that the viable LSP is
neutralino. Furthermore since $\mu$ now is free we can have either
bino-like or higgsino-like neutralino depending on the ratio of $\mu$ over
$m_{1/2}$. Thus the important coannihilation processes are: $\chi - \slept$ 
($\ell = e$, $\mu$, $\tau$), $\chi - \stop$, $\chi - \snu_\ell$,
and  $\chi -\chi^{\prime} - \chi^\pm$,  plus all combinations for the overlap
regions.

We impose in our analysis the constraints on the MSSM parameter space that are
provided by direct sparticle searches at LEP, including that on the lightest
chargino $\chi^\pm$: $m_{\chi^\pm} \gappeq$ 103.5 GeV, the selectron $\widetilde
e$: $m_{\tilde e}\gappeq$ 99 GeV,  and the Higgs mass: $m_H > 114.4$~GeV. We
also impose the constraint from measurements of $b\rightarrow s\gamma$, which
agree with the Standard Model calculation.  The latest value of the anomalous
magnetic moment of the muon, $a_\mu$, is also taken into account. However due
to the still lingering uncertainty in the Standard Model 
calculation of $a_\mu$, we do not impose this as an  absolute constraint on the
supersymmetric parameter space. These constraints and their corresponding
references are discussed in more detail in~\cite{NUHM}.

\section{Exploration of the Parameter Space}
\begin{figure}[t]
\begin{center}
\epsfxsize=15pc 
\epsfbox{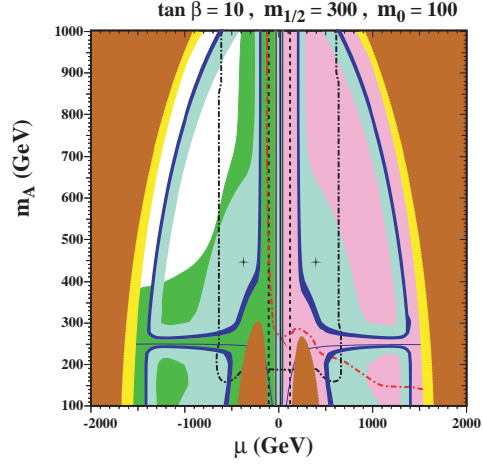} 
\end{center}
\caption{A $\mu - m_A$ plane in the
 NUHM model for $\tan \beta = 10$, $m_{1/2} = 300$ GeV, $m_0 = 100$ GeV and $A_0
 = 0$.  \label{fig:muma}}
\end{figure}

In Fig.~\ref{fig:muma} we plot a $\mu - m_A$ plane for $\tan \beta=10$,
$m_{1/2} = 300$ GeV, $m_0 =
100$ GeV and $A_0 = 0$. The shadings and lines are as follows. The
dark shaded regions have a charged sfermion lighter than the
neutralino, so these regions are excluded. Next to this regions at large $|\mu|$
are bands with light  shading which have sneutrino LSP.  
The $b \to s \gamma$ exclusion is presented by the
medium  shaded regions. The light  shaded areas are the
cosmologically preferred regions with $0.1\leq\ohsq\leq 0.3$. Also shown along
this region, shaded darkest, is the narrower range $0.094\leq\ohsq\leq
0.129$ from WMAP~\cite{wmap}.
The $\mu > 0$ region is shaded  to indicate that it satisfies the current
$2 \sigma$ range of muon anomalous magnetic moment, $11.5
\times 10^{-10} < \delta a_\mu < 56.3 \times 10^{-10}$. The  dot-dashed
line is
the contour $m_h = 114$~GeV. Regions on the left of this line are excluded.
The near-vertical  dashed lines are the contour
$m_{\chi^\pm} = 103.5$~GeV. Small $|\mu|$ region is excluded. 
The GUT
stability constraint is presented by the dark  dot-dashed lines. Only the
central region is allowed by this constraint. The near-horizontal solid 
lines at $m_A \sim 240$ GeV are for the heavy Higgs pole, $2 m_\chi = m_A$. 
The CMSSM cases are presented by the crosses, one for each sign of $\mu$. We see
that the NUHM allowed regions are much broader than the CMSSM ones, with
$\chi - \snu$ coannihilation regions at large $|\mu|$ and bino-higgsino
transition regions at small $|\mu|$. The GUT stability constraint, in this case,
excludes most of the $\chi - \snu$ coannihilation regions.

\begin{figure}[t]
\begin{center}
\epsfxsize=18pc 
\epsfbox{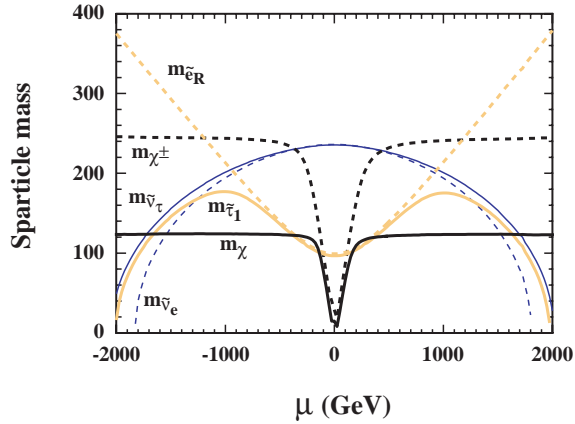} 
\end{center}
\caption{Sparticle mass spectrum as functions of
 $\mu$ for $\tan \beta = 10$,
 $m_{1/2} = 300$ GeV,  $m_0 = 100$ GeV, $A_0 = 0$ and 
 $m_A = 200$ GeV.  \label{fig:spect}}
\end{figure}

In Fig.~\ref{fig:spect}, we fix $m_A = 200$ GeV and plot the mass spectrum as
functions of $\mu$.  We see that for small $|\mu|$ we have higgsino $\chi$,
which is degenerate with the lightest chargino. Sneutrino LSP region is found
at large $|\mu|$, while $\stau$ LSP has relatively small $|\mu|$.

\section{Dark Matter Direct Detection Rates}
\begin{figure}[t]
\epsfxsize=14pc 
\epsfbox{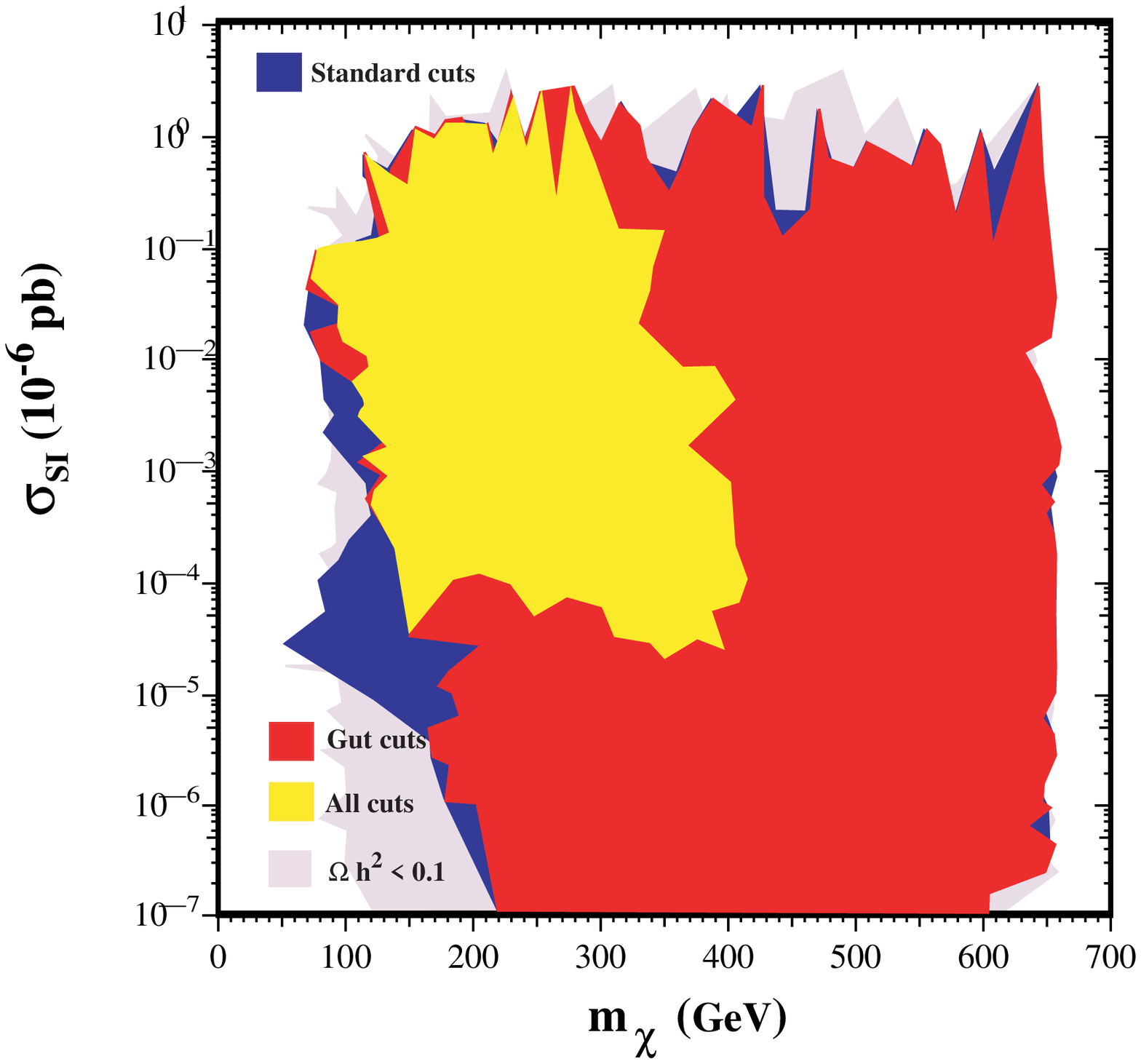}
\epsfxsize=14pc 
\epsfbox{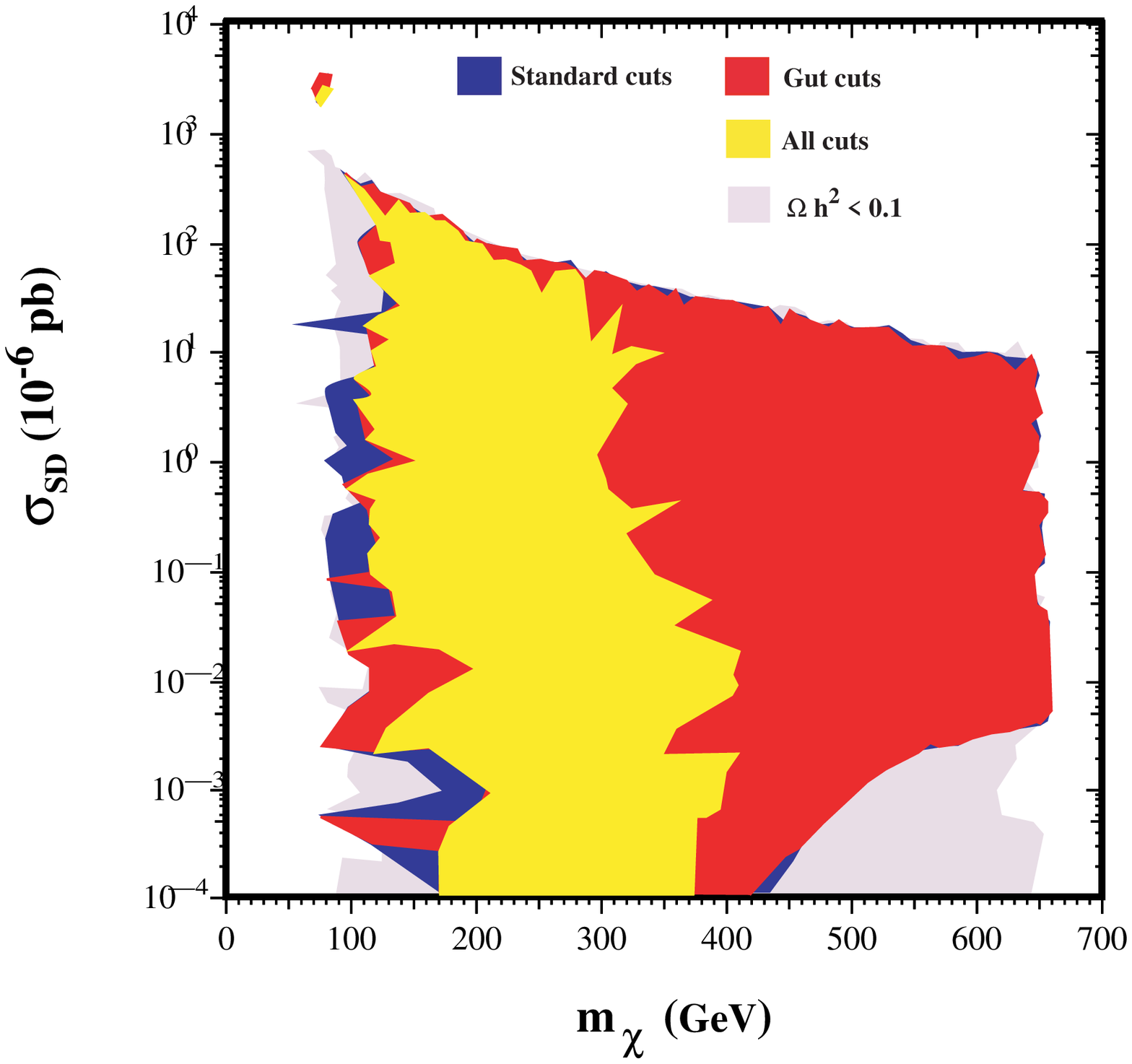}
\caption{Ranges of spin independent (scalar) and spin dependent cross section
in the NUHM model.  \label{fig:range}}
\end{figure}

We calculate the neutralino-proton elastic scattering cross section,
both spin dependent and spin independent parts. We cover the parameter
space by random scan and apply cuts from the experimental constraints. The
results for all
$\tan \beta$ are shown in Fig.~\ref{fig:range}. 
The light  shaded regions are what we get if we include constraint from
$a_\mu$.  
We find the range 
$10^{-10}\, {\rm pb} \lappeq \sigma_{SI} \lappeq 10^{-6}\, {\rm pb}$ and
$\sigma_{SD} \lappeq 10^{-3}$ pb. Compared with the CMSSM, the NUHM elastic
scattering could be up to two orders of magnitude large for similar neutralino
masses.

\section{Conclusion} 

Relaxing scalar mass universality for the  Higgs sector gives us
richer phenomenology compared to the CMSSM. There are  direct effects on the
sfermion mass
spectrum through RGE evolution. For example, here sneutrinos could
be degenerate with the neutralino LSP. The NUHM cosmological region consists of
the bulk,
the $H$ \& $A$ pole rapid annihilation,  the $\chi-\slept$ coannihilation, 
the $\chi-\snu$ coannihilation and
the Bino-Higgsino transition regions. (And also  $\chi -\stop$ coannihilation
region for $|A_0| \neq 0$ and large.) 
Bino-like  LSP is still preferred. Mostly
 Higgsino-like  LSP region is excluded either
by the  LEP constraints or because
$\chi -\chi^{\prime} -\chi^\pm$ coannihilation over-annihilates
the relic density. Note, however, that the second exclusion is based on the
assumption that dark matter consists only of neutralino LSP.

\section*{Acknowledgments}
This talk was based on work done in collaboration with J. Ellis, T. Falk, A.
Ferstl and K.A. Olive. This
work was supported in part by DOE grant  DE-FG02-94ER-40823.

\end{document}